\documentclass[%
 reprint,
https://www.overleaf.com/project/62a35d12098851462d494cde
 amsmath,amssymb,
 aps,
]{revtex4-2}

\usepackage{graphicx}
\usepackage{dcolumn}
\usepackage{bm}

\begin{document}
\preprint{APS/123-QED}

\title{Unimodular Gravity Traversable Wormholes} 

\author{A.S. Agrawal}
 \email{agrawalamar61@gmail.com}
\affiliation{Department of Mathematics, Birla Institute of Technology and Science-Pilani, Hyderabad Campus, Hyderabad-500078, India.}

\author{B. Mishra}
 \email{bivu@hyderabad.bits-pilani.ac.in}
\affiliation{Department of Mathematics, Birla Institute of Technology and Science-Pilani, Hyderabad Campus, Hyderabad-500078, India.}

\author{P.H.R.S. Moraes}
\email{moraes.phrs@gmail.com}
\affiliation{Universidade Federal do ABC (UFABC) - Centro de Ci\^encias Naturais e Humanas (CCNH) - Avenida dos Estados 5001, 09210-580, Santo Andr\'e, SP, Brazil}


\begin{abstract}

Wormholes are outstanding solutions of Einstein's General Relativity. They were worked out in the late 1980's by Morris and Thorne, who have figured out a recipe that wormholes must obey in order to be traversable, that is, safely crossed by travelers. A  remarkable feature is that General Relativity Theory wormholes must be filled by {\it exotic matter}, which Morris and Thorne define as matter satisfying $-p_r>\rho$, in which $p_r$ is the radial pressure and $\rho$ is the energy density of the wormhole. In the present article, we introduce, for the first time in the literature, traversable wormhole solutions of Einstein's Unimodular Gravity Theory. Unimodular Gravity was proposed by Einstein himself as the theory for which the field equations are the traceless portion of General Relativity field equations. Later, Weinberg has shown that this approach elegantly yields the solution of the infamous cosmological constant problem. The wormhole solutions here presented satisfy the metric conditions of ``traversability'' and remarkably evade the exotic matter condition, so we can affirm that Unimodular Gravity wormholes can be filled by ordinary matter.  

\end{abstract}

%

\maketitle

\section{Introduction}
\label{sec:i}

Traversable wormholes are outstanding and intriguing solutions of General Relativity field equations \cite{morris/1988}. Morris and Thorne in \cite{morris/1988} have proposed traversable wormholes as $i)$ a tool for teaching General Relativity, and $ii)$ an alternative for interstellar travel. 
Wormholes were already considered much before Morris and Thorne seminal article. One year after General Relativity appearance, Flamm recognized Schwarzschild solution as a representation of a wormhole \cite{flamm/1916}, which is referred to as {\it Schwarzschild wormhole}. The problem is that these wormholes cannot be traversable. The main reasons for the non-traversability of Schwarzschild wormholes are: the tidal gravitational forces at the wormhole throat are of the same order as in the black hole horizon and it would rapidly contract to a zero-radius circumference \cite{morris/1988,misner/2017}.  

Other types of wormhole solutions were proposed by Einstein and Rosen \cite{einstein/1935}, Weyl \cite{weyl/1949} and Wheeler \cite{wheeler/1962}, none of them being traversable, what reinforces (if there is any need) the importance of Morris and Thorne seminal article \cite{morris/1988}.

Morris and Thorne showed that in order to be traversable, the wormhole must obey a series of desired properties, suchlike \cite{morris/1988}: the solution must have a throat that connects two asymptotically flat regions of space-time, there should be no horizon and the tidal gravitational forces experienced by a traveler must be bearably small. These  properties will be revisited throughout the paper, together with further wormhole metric and energy-momentum tensor conditions. 

Several different forms to detect wormholes have been proposed, although with no success yet. The main possibility nowadays is gravitational lensing \cite{abe/2010}-\cite{godani/2021}. Particularly, in \cite{nandi/2017}, the similarities between wormholes and black holes ring-down gravitational waves and lensing observables were analyzed. Further possibilities on how to distinguish between wormholes and black holes can be seen in \cite{tsukamoto/2012,li/2014}. A detection method by use of images of wormholes surrounded by optically thin disk can be seen in \cite{ohgami/2015,wang/2020}.  Finally, a quite pragmatic possibility for observing a wormhole in the Galactic center was proposed in \cite{dai/2019}, through the study of the orbit of the stars around Sgr A*.

Another condition that, according to Morris and Thorne \cite{morris/1988}, traversable wormholes must obey is that they must be filled by a non-null amount of {\it exotic matter}, and by exotic matter they mean matter satisfying 

\begin{equation}\label{i1}
    -p_r>\rho,
\end{equation}
with the speed of light $c=1$ and such that $p_r$ is the radial pressure and $\rho$ is the matter-energy density of the wormhole.

This exotic matter condition has led many theoretical physicists to obtain wormhole solutions in gravitational theories departing from General Relativity with the purpose of investigating if the extra degrees of freedom presented in these theories would be able to allow the existence of traversable wormholes with no exotic matter. This outcome has been achieved through $f(R)$ gravity \cite{sotiriou/2010}, with $R$ being the curvature scalar, as one can check \cite{samanta/2019}, for instance.

Alternative gravity has been developed and applied with the main purpose of circumventing the cosmological constant problem \cite{weinberg/1989} in cosmology, with the physics of wormholes being another important field of application \cite{bronnikov/2003,bhadra/2005}. 

Here, it is important to mention that General Relativity presents some remarkable results when tested in different regimes, systems and situations \cite{will/1995}-\cite{turyshev/2008}. In this way, it is valuable to search for theories that minimally departure from General Relativity, which we might refer to as {\it theories of minimum extension} of General Relativity. Some theories of minimum extension examples  are the quadratic \cite{stelle/1977} and quartic gravity \cite{karasu/2016}.

It is the purpose of the present article to derive traversable wormhole solutions in a different theory of minimum extension of General Relativity, which is called today the {\it Einstein's Unimodular Gravity} \cite{einstein/1919}. Einstein has obtained Unimodular Gravity by taking the traceless portion of General Relativity field equations. Later, Weinberg showed that the aforementioned cosmological constant problem could be solved via Unimodular Gravity \cite{weinberg/1989}.

As a consequence, nowadays, Unimodular Gravity could be effectively seen as an alternative to the cosmological constant problem \cite{finkelstein/2001,jain/2012} and even to the problem of quantization of gravity \cite{yamashita/2020,smolin/2011}. Further applications of Unimodular gravity can be seen in cosmology \cite{gao/2014,garcia-aspeitia/2019} and stellar astrophysics  \cite{astorga-moreno/2019}.

\section{Einstein's Unimodular Gravity}
\label{sec:eug}

In 1919 Einstein started considering what today is known as Unimodular Gravity. In Unimodular Gravity, the metric determinant is kept fixed instead of being a dynamical variable as in General Relativity. Such a condition reduces the symmetry of the diffeomorphism group to the group of unimodular general coordinate transformations. Consequently, the equations governing the space-time dynamics are the traceless General Relativity field equations. 

Unimodular Gravity starts from the action \cite{garcia-aspeitia/2019}

\begin{equation}\label{eug1}
	S=\int d^4x\xi\left(\frac{R}{16\pi}+\mathcal{L}_m\right),
\end{equation}
with $\xi=\sqrt{-g}$ a fixed scalar density, $g$ the determinant of the metric $g_{\mu\nu}$, natural units are assumed and $\mathcal{L}_m$ is the matter lagrangian density.

By applying the variational principle in the above action yields the Unimodular Gravity field equations

\begin{equation}\label{eug2}
	R_{\mu\nu}-\frac{Rg_{\mu\nu}}{4}=8\pi\left(T_{\mu\nu}-\frac{Tg_{\mu\nu}}{4}\right),
\end{equation}
with $R_{\mu\nu}$ being the Ricci tensor, $T_{\mu\nu}$ the energy-momentum tensor and $T$ the trace of the energy-momentum tensor.

The above field equations can be rewritten as 

\begin{equation}\label{eug3}
	G_{\mu\nu}+\Lambda(R,T)g_{\mu\nu}=8\pi T_{\mu\nu},
\end{equation}
with $G_{\mu\nu}=R_{\mu\nu}-Rg_{\mu\nu}/2$ and 

\begin{equation}\label{x1}
   \Lambda(R,T)=\frac{R+8\pi T}{4}. 
\end{equation}
(\ref{eug3}) and (\ref{x1}) show that an important difference between General Relativity and Unimodular Gravity is in the nature of the cosmological constant. While in General Relativity it is a coupling constant put ``by hand'' in the lagrangian, in Unimodular Gravity it naturally emerges as a non-gravitating integration constant. In fact, the Unimodular Gravity field equations can therefore naturally contain an explication to the cosmic acceleration encoded in $\Lambda(R,T)$, with no cosmological constant problem. For $\Lambda(R,T)\rightarrow\Lambda$ (constant), Unimodular Gravity degenerates to General Relativity.

Note that a theory respecting (\ref{eug3}) resembles decaying vacuum energy models \cite{freese/1987,tong/2011}; however, while in such models the functional form in which $\Lambda$ depends on time is assumed {\it a priori} taking into account phenomenological reasons, in Unimodular Gravity the time-dependency of $\Lambda$ emerges naturally (recall (\ref{eug3}) and (\ref{x1})). 

\section{Traversable Wormholes: metric, energy-momentum tensor, traversability conditions and equation of state}
\label{sec:wh}

As it was explained in the Introduction, Morris and Thorne have worked out the conditions wormholes should satisfy in order to be traversable. We are going to visit these conditions in the present section. A deep discussion about them can be checked in \cite{morris/1988,visser/1996}.

The Morris-Thorne traversable wormhole is a static spherically symmetric space-time possessing two asymptotically flat regions. Recall that a fundamental property traversable wormholes must obey is the absence of horizons. 

The wormhole metric reads \cite{morris/1988}

\begin{equation}\label{wh1}
	ds^2=(e^\phi dt)^2-\frac{dr^2}{1-\frac{b}{r}}-r^2[d\theta^2+(\sin\theta d\varphi)^2],
\end{equation} 
in which $\phi=\phi(r)$ is the redshift function and $b=b(r)$ is the shape function, both unknowns. 

In order for the wormhole spatial geometry to tend to an appropriate asymptotically flat limit, one must require

\begin{eqnarray}
	\lim_{r\rightarrow\infty}\phi<\infty.\label{wh3}
\end{eqnarray}

At the wormhole throat, $r_0$, the traversable wormhole must obey

\begin{eqnarray}
	b(r_0)=r_0,\label{wh4}\\
	b'(r_0)\leq1,\label{wh5}
\end{eqnarray}
with $'\equiv d/dr$. We must also have

\begin{equation}\label{wh7}
	b'<\frac{b}{r}
\end{equation}
and away from the throat,

\begin{equation}\label{wh6}
	b<r.
\end{equation}

The wormhole energy-momentum tensor reads as \cite{morris/1988}

\begin{equation}\label{wh8}
	T^\mu_\nu=\texttt{diag}(\rho,-p_r,-p_t,-p_t),
\end{equation}
with $p_t$ being the tangential pressure. 

Morris and Thorne have shown that to be traversable, General Relativity wormholes must be filled by exotic matter, or matter satisfying (\ref{i1}).

Finally, the wormhole equation of state is still unknown and the eventual wormhole detection (recall References \cite{abe/2010}-\cite{dai/2019}) will certainly constraint it. It is common to see barotropic equations of state like

\begin{equation}\label{wh10}
	p_r=\alpha\rho,
\end{equation}
with constant $\alpha$, being invoked to describe wormholes \cite{garattini/2007}. Here, let us assume Eq.(\ref{wh10}) together with

\begin{equation}\label{wh11}
	p_t=\beta p_r,
\end{equation}
with constant $\beta$. The values of $\alpha$ can be restricted to $-1.5\leq\alpha\leq1$, so that we can describe, if necessary, different types of cosmological fluids, such as stiff matter \cite{carr/2010}, radiation \cite{bahamonde/2016}, dust \cite{kashargin/2020}, dark energy \cite{wang/2016} and phantom fluid \cite{cataldo/2013}. On the other hand, $\beta\neq1$ in order to guarantee anisotropy and $\neq0$ in order to avoid singularities. 

\section{Unimodular Gravity Traversable Wormholes}
\label{sec:ugtw}

The unimodular traversable wormholes will be obtained from the substitution of Eqs.(\ref{wh1}) and (\ref{wh8}) into Eq.(\ref{eug2}). By doing so, we obtain the following relations:

\begin{equation}\label{wh12}
\rho+3p_{r}-2p_{t}= \frac{2b-rb'}{4\pi r^{3}},
\end{equation}
\begin{equation}\label{wh13}
\rho-p_{r}+2p_{t}=\frac{-b}{4\pi r^3},
\end{equation}
\begin{equation}\label{wh14}
3\rho+p_{r}+2p_{t}=\frac{-b'}{4\pi r^2},
\end{equation}
where we have assumed $\phi(r)=\text{constant}$ in Eq.(\ref{wh1}). This can be done as it does not put in risk the asymptotically flat limit required in traversable wormholes, as one can check  (\ref{wh3}). A proper linear combination of Eqs.(\ref{wh12})-(\ref{wh14}) shows we have two independent equations with three unknowns.

It becomes essential to invoke an equation of state, such as (\ref{wh10}), to write

\begin{equation}
\rho=\frac{-b'r+b}{8\pi(\alpha+1)r^3}, \label{wh15}
\end{equation}
\begin{equation}
p_{t} =\frac{(-\alpha+1)b'r-(\alpha +3) b}{16\pi(\alpha+1)r^3}. \label{wh18}
\end{equation}

By taking $C$ as an integration constant, we can write the solution for the shape function as
\begin{equation}\label{b}
b(r)=Cr^\frac{\alpha(2\beta+1)+3}{\alpha(2\beta-1)+1}.
\end{equation}

Equation (\ref{wh4}) is straightforwardly respected when the constants $C=1$ and $r_0=1$, which will be assumed from now on.

In order to check the other metric conditions that must be satisfied in a traversable wormhole, we plot Figure \ref{fig1} below, in which the quantities $b(r)$, $b'(r)$ and $b(r)/r$ are depicted for $\beta=-1.1$. We observe that (\ref{wh6}) is respected for $r>r_0$ when $\alpha=0.8$. It is straightforward to check that condition (\ref{wh7}) is also satisfied. Remarkably, (\ref{wh5}) is also satisfied for $\beta=-1.1$, which can be verified by working out Equation (\ref{b}). 

\begin{figure}[h!]
\centering
\minipage{0.50\textwidth}
\includegraphics[width=7cm]{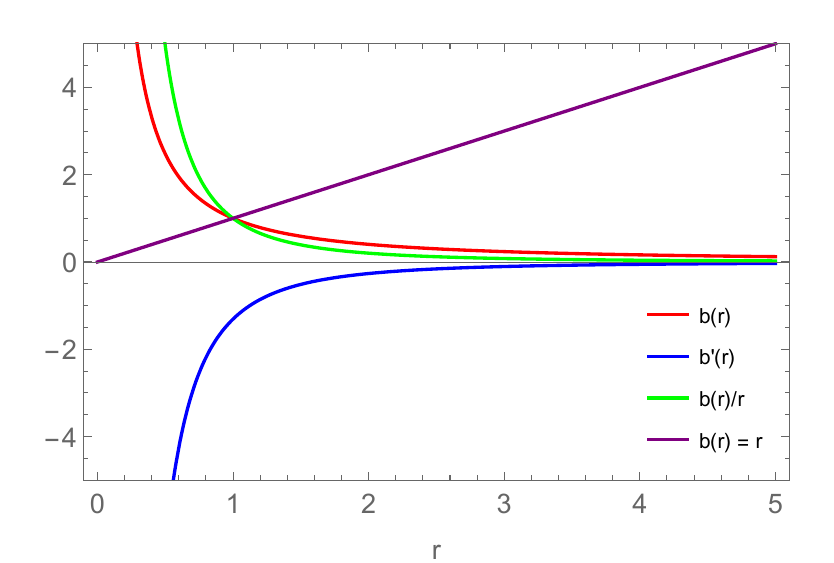}
\endminipage\hfill
\caption{$b(r)$, $b'(r)$ and $b(r)/r$ versus $r$ for  $\beta=-1.1$ and $\alpha=0.8$.}
\label{fig1}
\end{figure}

Now we will investigate the exotic matter issue for the Unimodular Gravity wormholes here obtained. As it was mentioned in Section 3, a General Relativity traversable wormhole must satisfy (\ref{i1}) and (\ref{wh3})-(\ref{wh6}), with the first one indicating the existence of exotic matter in the wormhole. We have just verified that  (\ref{wh3})-(\ref{wh6}) are satisfied for the Unimodular Gravity wormhole when $\beta=-1.1$ and now we need to check (\ref{i1}).

\begin{figure}[h!]
\centering
\minipage{0.50\textwidth}
\includegraphics[width=7cm]{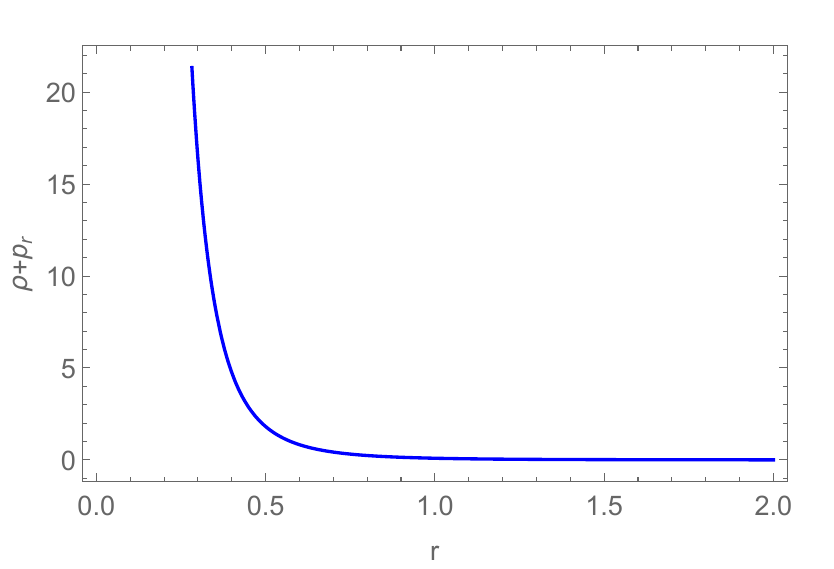}
\endminipage\hfill
\minipage{0.50\textwidth}
\includegraphics[width=7cm]{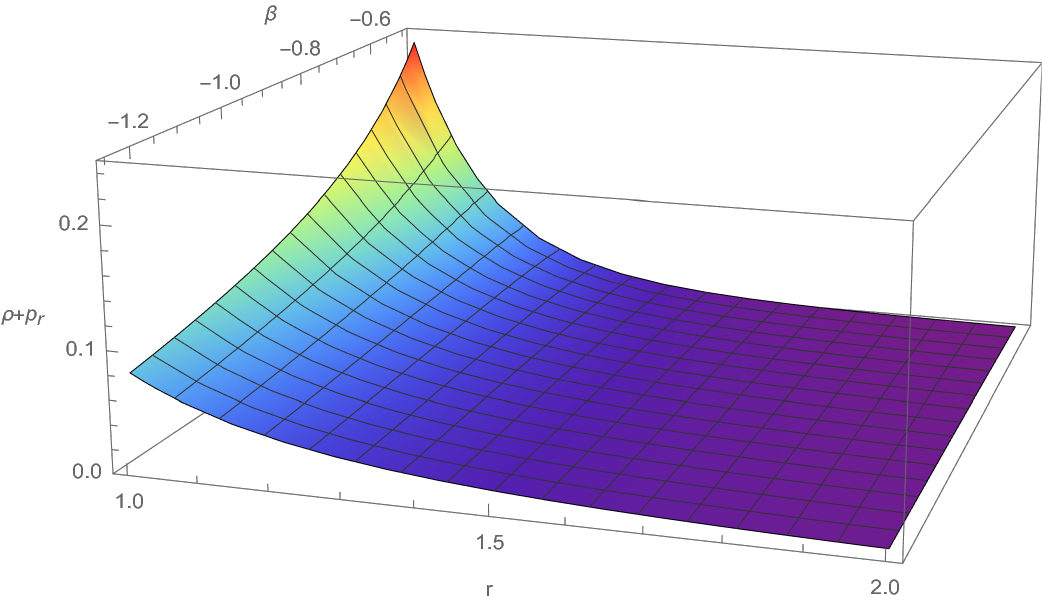}
\endminipage\hfill
\caption{$\rho +p_{r}$ versus $r$ for $\beta=-1.1$ (upper panel) and $\rho +p_{r}$ versus $r$ for the range $-1.25\leq\beta\leq -0.5$ (lower panel).}
\label{fig2}
\end{figure}

By plotting the quantity $\rho+p_r$ for $\beta=-1.1$ we obtain the results appearing in Fig.\ref{fig2} (upper panel).

Fig.\ref{fig2} is of extreme importance as it shows that $\rho+p_r\geq0$ for the Unimodular Gravity wormholes. As argued in \cite{morris/1988}, when matter inside wormholes respects (\ref{i1}), we can say the wormholes are filled by exotic matter. On the other hand, Fig.\ref{fig2} shows that Unimodular Gravity wormholes do not need to be filled by exotic matter in order to be traversable. 

In Fig.\ref{fig2} (lower panel) we show that this is not an exclusive property of $\beta=-1.1$ Unimodular Gravity wormholes, but it can be attained for a range of $\beta$ values.

\begin{figure}[h!]
\centering
\minipage{0.50\textwidth}
\includegraphics[width=7cm]{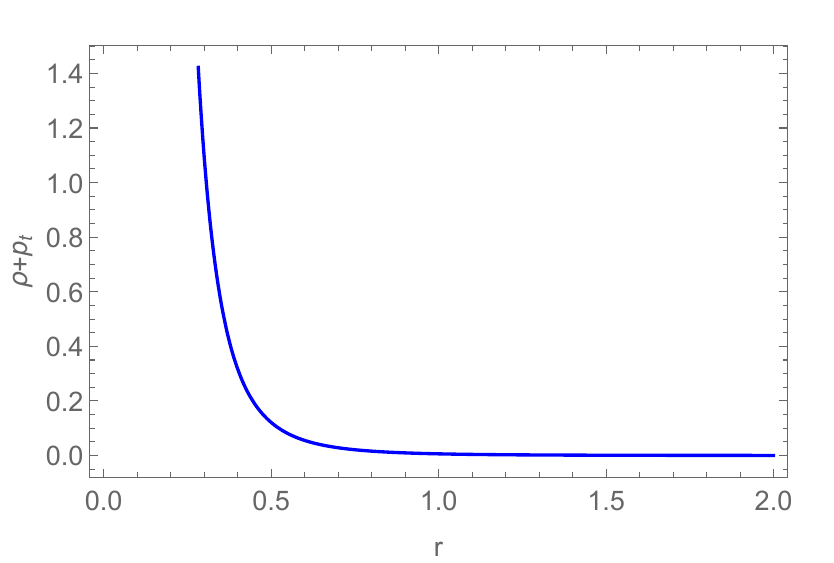}
\endminipage\hfill
\minipage{0.50\textwidth}
\includegraphics[width=7cm]{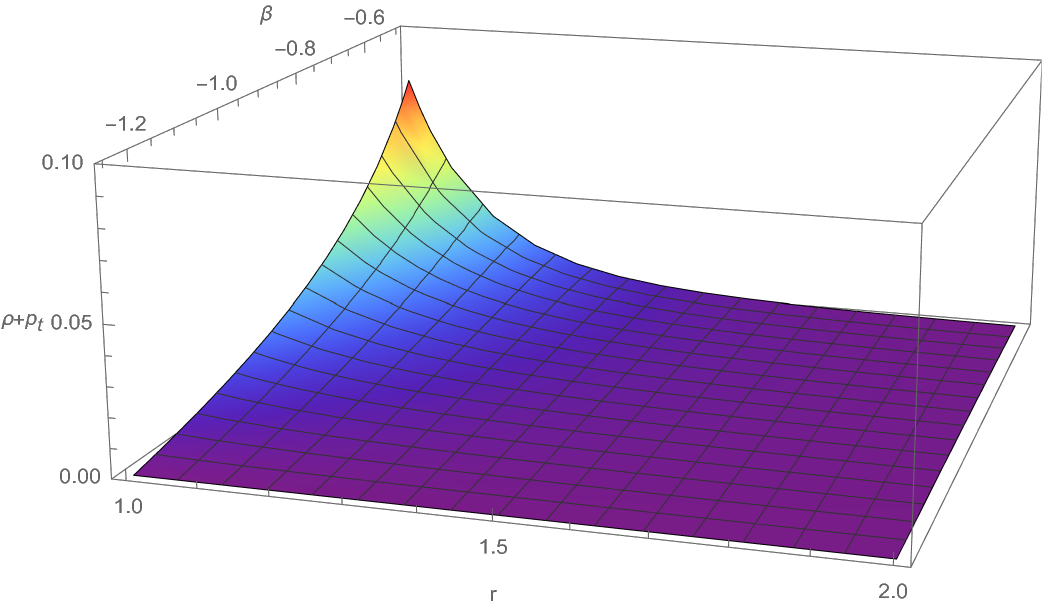}
\endminipage\hfill
\caption{$\rho +p_{t}$ versus $r$ for $\beta=-1.1$ (upper panel) and $\rho +p_{t}$ for the range $-1.25\leq\beta\leq -0.5$ (lower panel).}
\label{fig3}
\end{figure}

We can extend such an analysis for the $\rho+p_t$ case, which is seen in Fig.\ref{fig3}.

Figures \ref{fig2} and \ref{fig3} are nothing but the investigation of one of the {\it energy conditions} \cite{visser/1996} for the material content of the wormholes here obtained, namely the null energy condition. Physically, the null energy condition forces the local matter-energy density as measured by any timelike observer to be positive. 

The weak energy condition, which states that the matter content must satisfy the null energy condition together with $\rho\geq0$ \cite{visser/1996}, is also satisfied for the present Unimodular Gravity traversable wormholes.

\begin{figure}[h!]
\centering
\minipage{0.50\textwidth}
\includegraphics[width=7cm]{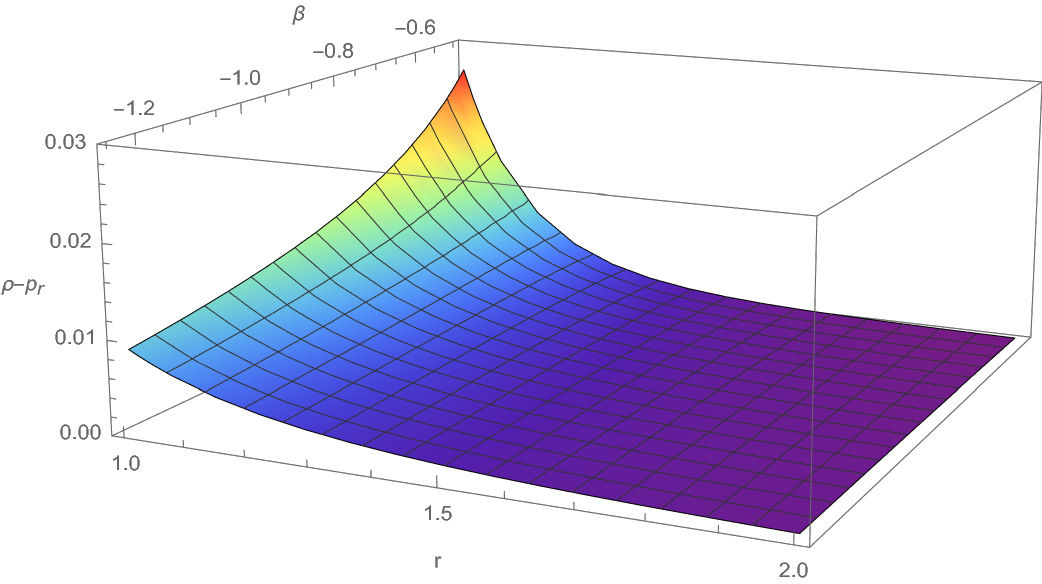}
\endminipage\hfill
\minipage{0.50\textwidth}
\includegraphics[width=9cm]{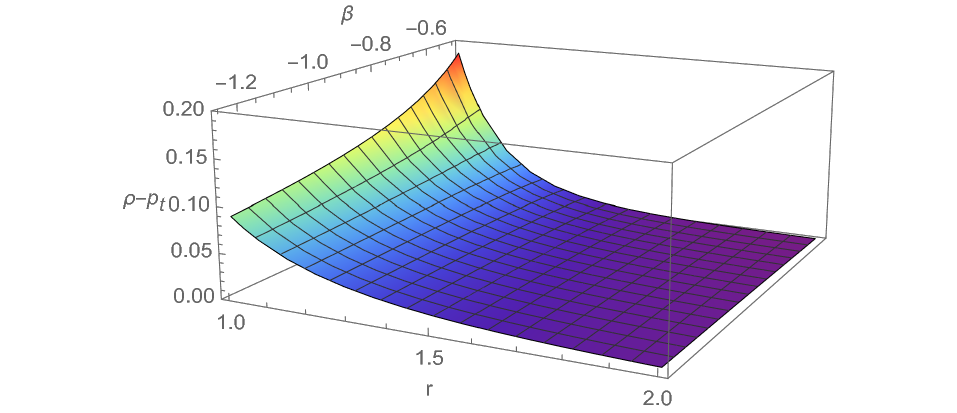}
\endminipage\hfill
\caption{$\rho -p_{r}$ (upper panel) and $\rho -p_{t}$ (lower panel) versus $r$  for the range $-1.25\leq\beta\leq -0.5$.}
\label{fig4}
\end{figure}

\begin{figure}[h!]
\centering
\minipage{0.50\textwidth}
\includegraphics[width=9cm]{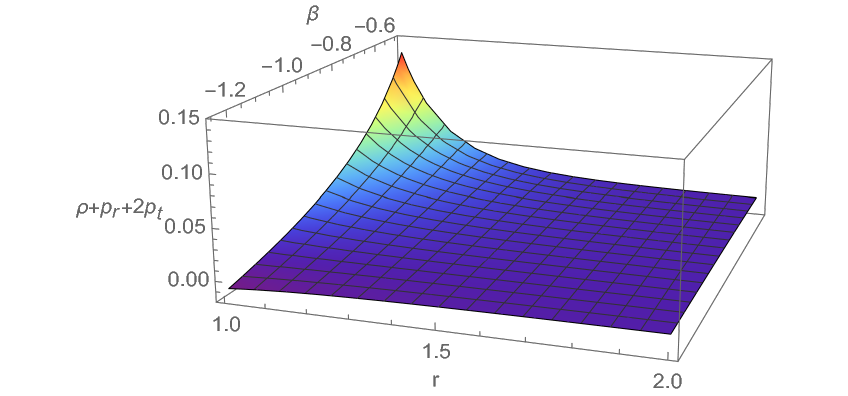}
\endminipage\hfill
\caption{$\rho +p_{r}+2p_{t}$ versus $r$ for the range $-1.25\leq\beta\leq -0.5$.}
\label{fig5}
\end{figure}

In Figures \ref{fig4} and \ref{fig5} below we plot the dominant energy condition and the strong energy condition, respectively. We can see that both are satisfied for a range of values of $\beta$ for the present Unimodular Gravity traversable wormholes, from which we can conclude that {\it all} the energy conditions are satisfied by Unimodular Gravity traversable wormholes.


\section{Unimodular Gravity Wormhole Structure} 

Here, we will describe the unimodular gravity wormhole space-time using the embedded diagram description and attempt to glean some details about the obtained shape function $b(r)$. Additionally, we determine the proper radial distance and demonstrate its length.

From the generic line element in Equation (6) concentrating on an equatorial plane, $\theta=\pi/2$, because we are working with spherically symmetric and static wormhole solutions, the solid angle element $d\Omega^{2}\equiv d\theta^{2}+(sin\theta d\varphi)^{2}$ reduces to
\begin{equation}
d\Omega^{2}= d\varphi^{2}.    
\end{equation}
Moreover, by fixing a constant time slide, i.e., $t=constant$, the line element (6) becomes 
\begin{equation}
ds^{2}=-\frac{dr^{2}}{1-\frac{b(r)}{r}}-(rd\varphi)^{2}.    
\end{equation}
It is practical to introduce cylindrical coordinates as a way to represent this equatorial plane as a surface embedded in a Euclidean space 
\begin{equation}
 ds^{2}=-dz^{2}-dr^{2}-(rd\varphi)^{2},   
\end{equation}
which is straightforwardly rewritten as
\begin{equation}
ds^{2}=-\left[1+\left(\frac{dz}{dr}\right)^{2}\right]dr^{2}-(rd\varphi)^{2}.  
\end{equation}
Now, comparing (22) and (24), we have
\begin{equation}
\frac{dz}{dr}=\pm \left[\frac{r}{b(r)}-1\right]^{-\frac{1}{2}},    
\end{equation}
where the embedded surface is defined by the equation $z=z(r)$. Interestingly, when $r\rightarrow +\infty$, the above formula results in
\begin{equation}
{\frac{dz}{dr}~ \vline }_{+\infty}=0.    
\end{equation}
This indicates that there are two asymptotically flat patches in the embedding diagram. Additionally, keep in mind that it is not possible to integrate Eq.(25) using Eq.(20) analytically. In order to depict the wormhole shape in the figure below, a numerical treatment has been done. The definition of a wormhole reveals that Eq.(25) is divergent at $r=r_{0}$, meaning the embedded surface is vertical there. The embedded surface in this example is the one in which we are assuming $\alpha=0.8$ and $\beta=-1.1$. The upper portion of the universe is represented by the red color in Figure 6 lower panel, while the lower portion is shown in green. The achievement of the wormhole hyper-surface involves rotating the curve $z(r)$ around a vertical axis. The radial distance is positive above the throat while is negative below the throat. It is also observed that far away from the throat the embedding surface becomes flat.

Morris and Thorne \cite{morris/1988} say that the proper radial distance must act properly everywhere, i.e., we must expect that $l(r)$ is finite everywhere in  space-time. To perform this, one must first be aware of the radial distance location of the stations (each portion of the universe). The same method used in \cite{morris/1988} can be used to obtain this information, namely that the factor $1-b(r)/r$ deviates from unity by no more than $1\%$. When this criterion is used and the numerical values $\{r_{0},\alpha, \beta\}=\{1,0.8,-1.1\}$[km] are taken into consideration, one obtains $\hat{r}=|r_{-}|=|r_{+}|\approx \pm 7.37$ Figure 7, so that
\begin{equation}
l(r)=\pm \int_{r_{0}^{+}}^{\hat{r}}\frac{dr}{\sqrt{1-\frac{b(r)}{r}}} \approx \pm 7.15   
\end{equation}
     
\begin{figure}
    \centering
    \minipage{0.50\textwidth}
    \includegraphics[width=6cm]{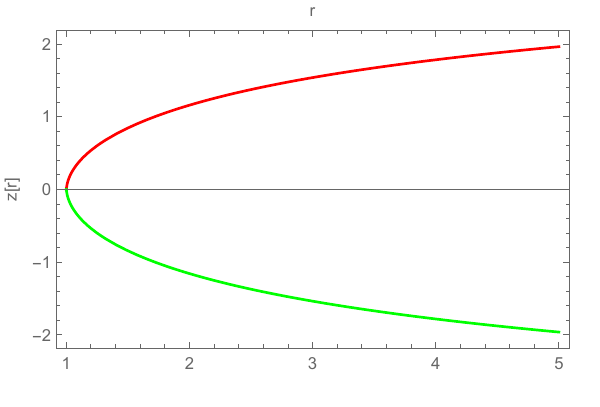}
    \endminipage\hfill
    \minipage{0.50\textwidth}
    \includegraphics[width=6cm]{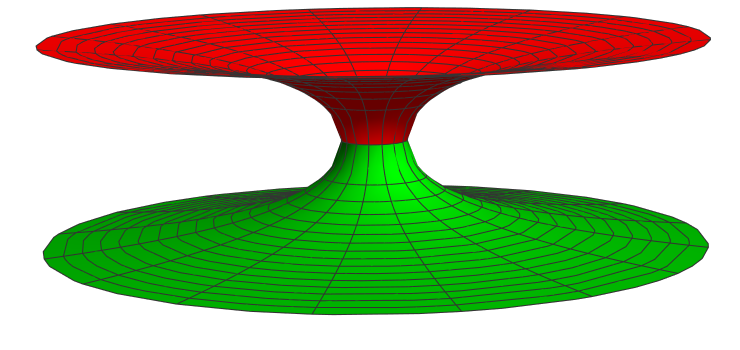}
    \endminipage\hfill
     \caption{The 2D $z(r)$ function (upper panel) was rotated around the vertical axis to create the 3D wormhole geometry (lower panel).}
    \label{fig6}
\end{figure}
\begin{figure}
    \centering
    \minipage{0.50\textwidth}
    \includegraphics[width=6cm]{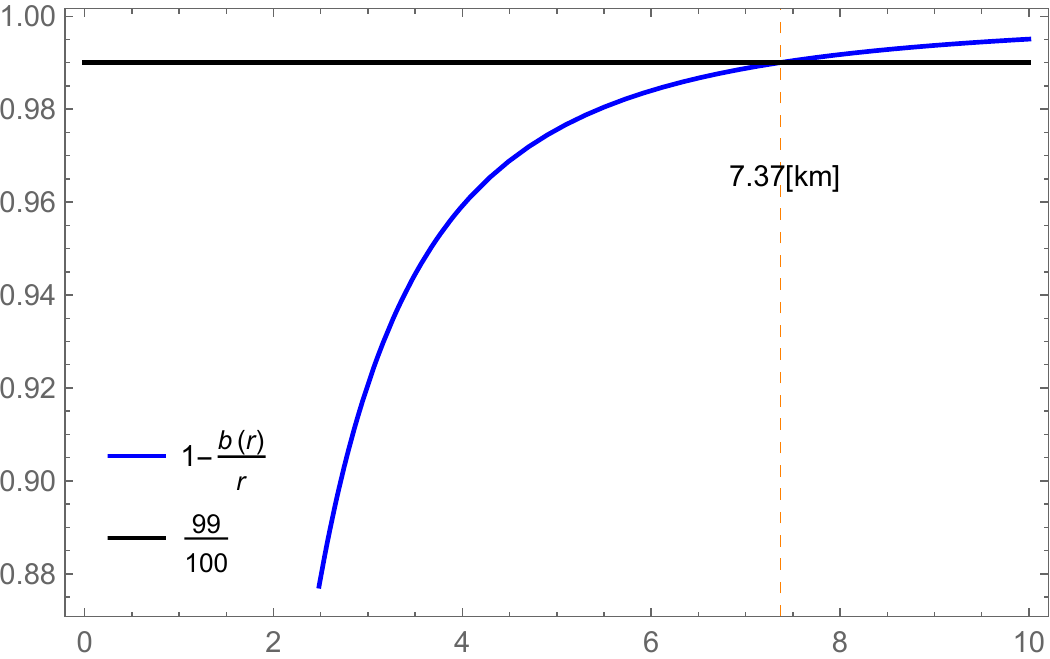}
    \endminipage
    \caption{The radial location $\hat{r}=|r_{\pm}|$ of spatial stations for values $\{r_{0},\alpha, \beta\}=\{1,0.8,-1.1\}$.}
    \label{fig7}
\end{figure}

The radial location of the spatial stations can be seen in the figure below.

\section{Discussion}
\label{sec:d}

Traversable wormholes are intriguing solutions of Einstein's General Theory of  Relativity \cite{morris/1988}. In the present article we have analysed, for the first time in the literature (to the best knowledge of the authors), wormhole solutions in the Einstein's Unimodular Theory of Gravity \cite{einstein/1919}. 

In 1919, Einstein started the consideration of what today is called Unimodular Gravity \cite{einstein/1919} by taking the traceless portion of General Relativity field equations, that can be retrieved from an integration constant, as it was shown in Section 2. Later, Weinberg showed that the problem related to the gravitational effects of vacuum quantum energy could be solved via Unimodular Gravity \cite{weinberg/1989}. By assuming Unimodular Gravity Theory field equations, the vacuum energy does not gravitate, differently from the General Relativity case. This does not determine an unique value for the effective cosmological constant, which solves the cosmological constant problem. Therefore, an important difference between General Relativity and Unimodular Gravity is in the nature of the cosmological constant; while in General Relativity it is a coupling constant in the Lagrangian, in Unimodular Gravity it, otherwise, elegantly arises naturally as an integration constant in the equations of motion. 

Today, Unimodular Theory of Gravity has been applied to different fields of research, yielding interesting and sometimes testable results, as one can check References \cite{finkelstein/2001}-\cite{astorga-moreno/2019}, for instance. Taking into account that General Relativity Theory has passed plenty of observational tests (check, for instance, \cite{will/1995}-\cite{turyshev/2008}, among many others), it is beneficial to invoke theories that minimally depart from it to approach some open issues in quantum, astrophysical or cosmological regimes. That is why we have chosen to work with Unimodular Gravity in the present article.

In General Relativity Theory, wormholes need to be filled by exotic matter. This kind of matter has not yet been observationally/experimentally probed and according to General Relativity, it would cause some kind of anti-gravitational effect to keep the wormhole throat opened  (avoiding its collapse). Here, we wanted to investigate if Unimodular Gravity could provide this effect purely from its gravity description with no need to impose exotic matter existence. 

This is what we obtained. In Fig.\ref{fig2} we see that it is possible to attain wormholes disrespecting (\ref{i1}) in Unimodular Gravity. In other words, it is possible to obtain wormholes free of exotic matter in this gravity theory. Remarkably, the metric conditions regarding the asymptotically flat limit (\ref{wh3}) and traversability (\ref{wh4})-(\ref{wh6}) are all respected, at least for a range of values of the free parameter $\beta$. We can also see from the three-dimensional plots in Figures \ref{fig2} and \ref{fig3} that the null energy condition is satisfied in the wormhole throat for a range of values of $\beta$.  When (\ref{i1}) is respected we have the null energy condition violation. In fact, the null energy condition is also written in terms of $\rho+p_t$, which is why we plotted Fig.\ref{fig3}, that together with Fig.\ref{fig2}, points to the null energy condition respectability.

Therefore, the Unimodular Gravity Theory wormholes here presented remarkably respect the null energy condition. Morris and Thorne qualified the null energy condition violation as ``troublesome'' since a traveler moving sufficiently fast through a null energy condition violating wormhole would measure a negative matter-energy density \cite{morris/1988}.

 Also remarkably, all energy conditions are satisfied for Unimodular Gravity traversable wormholes. Such a feature is scarcely seen in the present literature (check, for instance, \cite{harko/2013}-\cite{forghani/2020}).

With the purpose of investigating the metric conditions, we have plotted Fig.\ref{fig1}. To depict $b(r)$, $b'(r)$ and $b(r)/r$ we have chosen some particular values of $\beta$. For $\beta=-1.1$, all the metric conditions of traversability are fulfilled. Remarkably, (\ref{wh6}) is respected exactly in the $r>r_0$ range, as required \cite{visser/1996}. \\

{\bf Acknowledgements}\\

ASA acknowledges the financial support provided by University Grants Commission (UGC) through Senior Research Fellowship (File No. 16-9 (June 2017)/2018 (NET/CSIR)), to carry out the research work. BM thanks IUCAA, Pune, India for providing academic support through visiting associateship program. We are thankful to the anonymous reviewer for the comments and suggestions to improve the quality of the paper.


\begin{thebibliography}{00}

\bibitem{morris/1988} M.S. Morris and K.S. Thorne, Amer. J. Phys. 56 (1988) 395. 
\bibitem{flamm/1916} L. Flamm, Phys. Z. 17 (1916) 448.
\bibitem{misner/2017} C.W. Misner et al., {\it Gravitation} (Princeton University Press, New Jersey, USA, 2017).
\bibitem{einstein/1935} A. Einstein and N. Rosen, Phys. Rev. 48 (1935) 73.
\bibitem{weyl/1949} H. Weyl, {\it Philosophy of Mathematics and Natural Science} (Princeton University Press, New Jersey, USA, 1949).
\bibitem{wheeler/1962} J.A. Wheeler, {\it Geometrodynamics} (Academic, New York, USA, 1962).
\bibitem{abe/2010} F. Abe, Astrophys. J. 725 (2010) 787.
\bibitem{toki/2011} Y. Toki et al., Astrophys. J. 740 (2011) 121.
\bibitem{godani/2021} N. Godani and G.C. Samanta, Ann. Phys. 429 (2021) 168460.
\bibitem{nandi/2017} K.K. Nandi et al., Phys. Rev. D 95 (2017) 104011.
\bibitem{tsukamoto/2012} N. Tsukamoto et al., Phys. Rev. D 86 (2012) 104062.
\bibitem{li/2014} Z. Li and C. Bambi, Phys. Rev. D 90 (2014) 024071.
\bibitem{ohgami/2015} T. Ohgami and N. Sakai, Phys. Rev. D 91 (2015) 124020.
\bibitem{wang/2020} X. Wang et al., Phys. Lett. B 811 (2020) 135930.
\bibitem{dai/2019} D.-C. Dai and D. Stojkovic, Phys. Rev. D 100 (2019) 083513.
\bibitem{sotiriou/2010} T.P. Sotiriou and V. Faraoni, Rev. Mod. Phys. 82 (2010) 451.
\bibitem{samanta/2019} G.C. Samanta and N. Godani, Mod. Phys. Lett. A 34 (2019) 1950224.
\bibitem{weinberg/1989} S. Weinberg, Rev. Mod. Phys. 61 (1989) 1.
\bibitem{bronnikov/2003} K.A. Bronnikov and S.-W. Kim, Phys. Rev. D 67 (2003) 064027.
\bibitem{bhadra/2005} A. Bhadra and K. Sarkar, Mod. Phys. Lett. A 20 (2005) 1831.
\bibitem{will/1995} C.M. Will, {\it Was Einstein Right? Putting General Relativity to the Test} (Oxford University Press, Oxford, UK, 1995).
\bibitem{will/1984} C.M. Will, Phys. Rep. 113 (1984) 345.
\bibitem{will/2006} C.M. Will, Ann. Phys. 15 (2006) 19.
\bibitem{will/2014} C.M. Will, Liv. Rev. Rel. 17 (2014) 117.
\bibitem{turyshev/2008} S.G. Turyshev, Ann. Rev. Nucl. Part. Sci. 58 (2008) 207.
\bibitem{stelle/1977} K.S. Stelle, Phys. Rev. D 16 (1977) 953.
\bibitem{karasu/2016} A. Karasu et al., Phys. Rev. D 93 (2016) 084040.
\bibitem{einstein/1919} A. Einstein, Sitzungs. Preuss. Akad. Wiss. (1919) 142.
\bibitem{finkelstein/2001} D.R. Finkelstein et al., J. Math. Phys. 42 (2001) 340.
\bibitem{jain/2012} P. Jain et al., J. Cosm. Astrop. Phys. 11 (2012) 003.
\bibitem{yamashita/2020} S. Yamashita, Phys. Rev. D 101 (2020) 086007.
\bibitem{smolin/2011} L. Smolin, Phys. Rev. D 84 (2011) 044047.
\bibitem{gao/2014} C. Gao et al., J. Cosm. Astrop. Phys. 09 (2014) 021.
\bibitem{garcia-aspeitia/2019} M.A. Garc\'ia-Aspeitia et al., Phys. Rev. D 99 (2019) 123525.
\bibitem{astorga-moreno/2019} J.A. Astorga-Moreno et al., J. Cosm. Astrop. Phys. 09 (2019) 005.
\bibitem{freese/1987} K. Freese et al., Nucl. Phys. B 287 (1987) 797.
\bibitem{tong/2011} M. Tong and H. Noh, Eur. Phys. J. C 71 (2011) 1586.
\bibitem{visser/1996} M. Visser, {\it Lorentzian Wormholes: from Einstein to Hawking} (AIP Press, New York, USA, 1996).
\bibitem{garattini/2007} R. Garattini, Class. Quant. Grav. 24 (2007) 1189.
\bibitem{carr/2010} B.J. Carr et al., Class. Quant. Grav. 27 (2010) 183101.
\bibitem{bahamonde/2016} S. Bahamonde et al., Phys. Rev. D 94 (2016) 044041.
\bibitem{kashargin/2020} P.E. Kashargin and S.V. Sushkov, Universe 6 (2020) 186.
\bibitem{wang/2016} D. Wang and X.-H. Meng, Eur. Phys. J. C 76 (2016) 484.
\bibitem{cataldo/2013} M. Cataldo and P. Meza, Phys. Rev. D 87 (2013) 064012.
\bibitem{harko/2013} T. Harko et al., Phys. Rev. D 87 (2013) 067504.
\bibitem{miranda/2019} G. Miranda et al., Phys. Rev. D 99 (2019) 124045.
\bibitem{forghani/2020} S.D. Forghani et al., Phys. Lett. B 804 (2020) 135374.
\end{thebibliography}
\end{document}